\documentclass[runningheads]{llncs}

\usepackage{graphicx}
\usepackage{array}
\newcolumntype{?}{!{\vrule width 1pt}}
\usepackage{amsmath}
\usepackage{amssymb}
\usepackage{multirow}
\usepackage[colorinlistoftodos]{todonotes}
\usepackage[colorlinks=true, allcolors=blue]{hyperref}
\usepackage{tikzsymbols}
\usepackage[linesnumbered,ruled,vlined]{algorithm2e}

\usepackage{stfloats}
\usepackage{boldline}
\usepackage[colorinlistoftodos]{todonotes}
\usepackage{comment}
\usepackage{cite}
\usepackage{color}
\usepackage{graphicx}
\usepackage{subcaption}
\usepackage{caption,stackengine}

\newcommand{\x}{\mathbf{x}}
\newcommand{\y}{\mathbf{y}}
\newcommand{\z}{\mathbf{z}}

\SetKwInput{KwInput}{Input}                
\SetKwInput{KwOutput}{Output}

\begin{document}
\title{Variational Knowledge Distillation for Disease Classification in Chest X-Rays}

\author{Tom van Sonsbeek\inst{1}  \and
Xiantong Zhen\inst{1,2} \and
Marcel Worring\inst{1} \and Ling Shao\inst{2} }
\authorrunning{T van Sonsbeek et al.}
\institute{University of Amsterdam, The Netherlands \and Inception Institute of Artificial Intelligence, U.A.E\\
\email{\{t.j.vansonsbeek, x.zhen, m.worring\}@uva.nl},
\email{ling.shao@ieee.org}}\looseness=-1
%

% \authorrunning{Tom van Sonsbeek \textit{et al.}}

% If there are more than two authors, '\textit{et al.}' is used.
%
% \institute{University of Amsterdam\\
% \email{\{t.j.vansonsbeek,x.zhen, m.worring\}@uva.nl}
% If there are more than two authors, '\textit{et al.}' is used.
%

%
\maketitle              % typeset the header of the contribution
\begin{abstract}

% In this paper we propose a method which shows how to distill information from Electronic Health Records (EHR) into image-based classification methods in a probabilistic inference framework. Image-based methods in medical imaging attempt to replicate certain clinician analyses, while solely relying on the image. Clinicians can benefit from contextualized patient information from EHR to reach their conclusion. Only few attempts have been made to include this contextual patient information in image-based methods. 
% EHR being dependant on clinician input a restive reason why EHR are not widely adopted to enhance tasks in medical image analysis. When EHR are needed as input a medical imaging can not always be automated anymore. This method aims to solve this by utilising EHR in training time of models, without the need for their presence during test time. To achieve this we make use of conditional variational models in which stochastic image and EHR distributions are conditioned on each other.
% We show show the benefit of conditional variational inference on current public datatsets which contain paired EHR and medical images. 

Disease classification relying solely on imaging data attracts great interest in medical image analysis. Current models could be further improved, however, by also employing Electronic Health Records (EHRs), which contain rich information on patients and findings from clinicians. It is challenging to incorporate this information into disease classification due to the high reliance on clinician input in EHRs, limiting the possibility for automated diagnosis. In this paper, we propose \textit{variational knowledge distillation} (VKD), which is a new probabilistic inference framework for disease classification based on X-rays that leverages knowledge from EHRs. Specifically, we introduce a conditional latent variable model, where we infer the latent representation of the X-ray image with the variational posterior conditioning on the associated EHR text. By doing so, the model acquires the ability to extract the visual features relevant to the disease during learning and can therefore perform more accurate classification for unseen patients at inference based solely on their X-ray scans. We demonstrate the effectiveness of our method on three public benchmark datasets with paired X-ray images and EHRs. The results show that the proposed variational knowledge distillation can consistently improve the performance of medical image classification and significantly surpasses current methods.
\looseness=-1
\end{abstract}

\section{Introduction}

Advances in deep learning for medical imaging have been shown to perform on par or better than clinicians on an increasing number of tasks\cite{liu2019comparison}. The expansion of data and computational resources has played a large role in this. In fact, while deep learning models and clinicians may seem very different at first, their underlying prediction process is similar, as they both acquire experience through data. However, clinicians currently have an advantage; their decision making is not only based on medical images. In addition to their own knowledge and experience, information on the patient can also provide important guidance when making a diagnosis. Thus, there is an opportunity for deep learning methods to be even further improved if they could also incorporate this information.

EHRs contain rich information about the patients, which could be explored for disease classification based on X-ray scans. Besides important patient information, e.g., disease history, sex and reason for hospital admission, they record observations and findings that are usually provided by clinicians from reading the scans in combination with their professional knowledge and clinical experiences. It has been demonstrated that (longitudinal) EHR data can be used as a diagnosis predictor \cite{Xiao2018, Tobore2019, Harerimana2019, Cai2019}. Thus it would be greatly helpful to if we can leverage EHRs to support clinicians by improving the performance of various automated medical imaging tasks.

However, it is a challenging problem to incorporate information from EHRs into medical image analysis due to several reasons. Firstly, representing EHR in models is complicated by the large variety in content, structure, language, noise, random errors, and sparseness \cite{Cai2019,Tobore2019}. Secondly, there are privacy concerns in using medical images associated with (longitudinal) EHR data, as their combined use limits the extent of possible anonymization~\cite{WEISKOPF2013830}. However the major limiting factor is that combining visual and textual modalities adds complexity, because it requires methods that span both vision and language processing fields.   %The last two factors will be discussed in more detail. 

From a clinical point of view, the usage of EHR data available during testing or model deployment should be approached with caution, because EHR data is not always available coupled to the patient at the time of diagnosis. It is important to keep in mind that tasks performed on medical data are only relevant in a clinical setting. Requiring EHR data as input for a model would prevent this model from being completely automated, because the EHR still needs to be created by a clinician. However, this is not a problem during training when access to large databases of medical images and EHRs is possible. Therefore, to effectively utilize EHRs in combination with medical images, they should be optimally utilized during training time, with minimum reliance on them during testing. In a clinical setting, this would make most sense, because we would like the model to assist clinicians rather than relying on them.

It is particularly appealing to leverage information in EHRs for disease classification of X-ray scans. This is because chest X-rays are one of the most common scans in clinical routines due to their ease of acquiring and low cost, offering an effective and efficient tool for screening various diseases. A consequence of this is a large quantity of scans, the bulk of which will fall under a frequent set of diagnoses. Both the potential usefulness of EHRs and importance of automated diagnosis in clinical setting make X-rays an excellent application domain. 

In this work, we tackle the challenging scenario where the EHRs are only available in the learning stage but not at inference time. We propose variational knowledge distillation, a new probabilistic inference framework to leverage knowledge in EHRs for medical image classification. \looseness=-1
%By transforming both images and EHR to probabilistic distributions we can condition images on EHR, such that certain indicative information inside the image is 'unlocked'.
We make three-fold contributions: \textit{i}) We propose the first probabilistic inference framework for joint learning from EHRs and medical images, which enables us to explore multi-modal data during training and perform disease classification relying only on images during testing. \textit{ii}) We introduce variational knowledge distillation, which enables the model to extract visual features relevant to disease by transferring useful information from EHRs to images. \textit{iii}) We demonstrate the effectiveness of the proposed method in X-ray disease classification and achieve consistently better performance than counterpart methods. \looseness=-1

% Therefore we propose to condition medical images from a large database of images and EHR to the EHR. This model should ‘unlock’ certain indicative information present in the image to benefit downstream tasks. This tasks is performed using a variational architecture. This architecture encodes the input data in a probabilistic distribution from which a latent representation of the input data can be sampled. EHR information is used to encode the 

% In the currently biggest public vision-language dataset in medical imaging, MIMIC-CXR (~240 000 image-EHR pairs), each pair can be classified in 14 classes. \tom{this belongs somewhere else} The class is determined based on the content of the EHR. Consequently, the EHR is a 1-to-1 predictor for the class label given to the image-EHR pair. (verbatim presence of class label in EHR occurs in ~80%) Recording a classification result based on this data is therefore not a useful task. 

\section{Related Work}

In recent years there has been an increase in methods exploring automated diagnosis from radiology images. This can be linked to the increasing availability of public chest X-ray datasets, such as ChestX-ray14 \cite{wang2017chest8}, CheXpert \cite{irvin2019chexpert}, OpenI  \cite{openi} and MIMIC-CXR \cite{johnson2019mimic}, where the latter two also contain associated EHR. \looseness=-1

The most notable image-based approach for chest X-ray classification is ChexNet \cite{rajpurkar2017chexnet}. Rajpurkar \textit{et al.} showed that diagnosis using a deep architecture based on DensetNet-121\cite{huang2017densely} can exceed radiologist performance. Wang \textit{et al.} \cite{wang2017chest8} also reached high performance using pre-trained convolutional neural network (CNN). Recently, Chen \textit{et al.} \cite{chen2020label} introduced a graph based model which exceeds the performance of the prior methods in this classification task. \looseness=-1

Current multi-modal approaches for chest X-ray classification rely on EHR inputs during both training and testing. A common denominator in these methods is that image and EHR features are joined through an attention mechanism. Nunes \textit{et al.}~\cite{Nunes2019} proposed a method which requires a chest X-ray and its associated EHR to generate a diagnosis. Wang \textit{et al.}~\cite{wang2018tienet} require a similar input but use an auxiliary EHR generation task in an end-to-end CNN-recurrent neural network~(RNN) architecture to improve classification. Related to this, Xue \textit{et al.}~\cite{xue2019improved} generate EHRs to enhanced image-based classification. No EHR input is required or used during both training and testing. Where our approach uses both image and EHR during training, but only images during testing, their approach only requires images in both training and testing.

Recent advances in the general non-medical vision-language field have been accelerated by the emergence of contextual Transformer \cite{vaswani2017attention} based language models such as BERT \cite{devlin2018bert}. Moreover in visual-question-answering (VQA), models such as LXMERT \cite{tan2019lxmert}, VL-BERT \cite{su2019vl}, VILBERT \cite{lu2019vilbert} and Uniter \cite{chen2019uniter} vastly outperform traditional state-of-the-art models. The Transformer architecture has proven to be highly effective in multi-modal settings.  Recently, Li \textit{et al.} \cite{li2020medtransform} showed how these vision-language models can be applied to the medical domain. Specifically, they showed that Transformer-based vision-language models result in high performance on medical datasets containing chest X-ray images and paired EHRs, requiring both modalities as input during training and testing.

% A more acceptable way of using medical images would be to 

% Variational Auto-Encoder (VAE) and Variational Information Bottleneck (VIB) are techniques in which data input is transformed to a probabilistic distribution based on which a latent space representation of said data input can be sampled. This latent space representation is used in downstream tasks or used to reconstruct the given data input for VIB and VAE respectively. 

\section{Methodology}
We formulate the disease classification from medical images as a conditional variational inference problem. We introduce a conditional latent variable model that infers the latent representations of X-Ray images. The knowledge is transferred from the EHR to X-rays scans by making the variational posterior conditioned on the associated EHR text. The model learns the ability to extract visual features that are relevant to the disease guided by the associated EHR in the learning stage. At inference time it is able to make accurate predictions relying solely on X-ray scans. We start with preliminaries on variational auto-encoders \cite{kingma2013auto,rezende2014stochastic}, based on which we derive our probabilistic modeling of disease classification on X-rays and variational knowledge distillation from EHRs.

%Our goal is to transfer knowledge encoded in EHR to medical images by conditioning the EHR. 
%The proposed architecture consists of two branches, the image-branch $I$ and the EHR-branch $R$ \autoref{fig:overview}. Moreover, it consists of three parts: First a modality specific extraction of features. Secondly a transformation of these features into Gaussian distributions. A latent space representation $z$ of the features can be sampled from these probabilistic distributions. In train time we can condition the distribution of the EHR branch, called the recognition network $q_\phi(z_R\|R)$, to the distribution of the image branch, the prior network $p_\theta(z_I\|I)$. Lastly we have stochastic feed-forward inference on the image branch by passing samplings from the latent space distributions of the image branch through a classification head. For this last step no EHR input is needed during inference time. 

\subsection{Preliminaries}
%I would like to talk a bit about variational auto-encoder/variational information bottleneck in this subsection.
The variational auto-encoder (VAE) \cite{kingma2013auto,rezende2014stochastic} is a powerful generative model that combines graphical models and deep learning. Given an input $\mathbf{x}$ from a data distribution $p(\mathbf{x})$, we aim to find its representation $\mathbf{z}$ in a latent space, from which we can generate new images that are similar to $\mathbf{x}$. The objective of the VAE is to maximize what is called the evidence lower bound (ELBO), as follows:
\begin{equation}
    \mathcal{L}_{\rm{VAE}} = \mathbb{E}[\log p(\mathbf{x}|\mathbf{z})] - D_{\rm{KL}}[q(\mathbf{z}|\mathbf{x})||p(\mathbf{z})],
\end{equation}
where $q(\mathbf{z}|\mathbf{x})$ is the variational posterior for approximating the exact posterior $p(\mathbf{z}|\mathbf{x})$ and $p(\mathbf{z})$ is the prior distribution over $\mathbf{z}$, which is usually set to an isotropic Gaussian distribution $\mathcal{N}(0,I)$. The VAE offers an effective probabilistic inference framework to learn latent representations in a unsupervised way, which we explore for the supervised, disease classification task by introducing conditioning into the probabilistic framework.
 
%\paragraph{Variational Information Bottleneck}

\subsection{Disease Classification by Conditional Variational Inference}
Since disease classification based on X-rays is a supervised learning problem, we resort to conditional variational inference, which has shown great effectiveness in structure prediction tasks~\cite{sohn2015learning}. Given an input X-ray image $\x_I$ associated with its class label $\y$, we introduce the latent variable $\z_I$ as the representation of $\x_I$. From a probabilistic perspective, predicting of the class label $\y$ amounts to maximizing the following conditional log-likelihood:
\begin{equation}
    \log p(\y|\x_I) = \log \int p(\y|\x_I,\z_I) p(\z_I|\x_I)d\z_I,
    \label{cll}
\end{equation}
where $p(\z_I|\x_I)$ is the conditional prior over the latent representation $\z_I$ (See Fig.~\ref{fig:overview}). To find the posterior $p(\z_I|\x_I,\y)$ over $\z_I$, we usually resort to a variational distribution $q(\z_I)$ by minimizing the Kullback-Leibler (KL) divergence
\begin{equation}
    D_{\rm{KL}} \big[ q(\z_I)||p(\z_I|\x_I,\y)\big].
    \label{kl}
\end{equation}
By applying Bayes' rule, we obtain 
\begin{equation}
    \mathcal{L}_{\rm{CVI}} = \mathbb{E}\big[\log p(\y|\x_I,\z_I)\big] - D_{\rm{KL}}\big[q(\z_I)||p(\z_I|\x_I)\big],
\end{equation}
which is the ELBO of the conditionally predictive log-likelihood in Eq.~(\ref{cll}) and can be directly maximized to learn the model parameters. Note that maximizing the ELBO is equivalent to minimizing the KL divergence in Eq.~(\ref{kl}). Actually, we are free to design the variational posterior $q(\z)$. In this work, we incorporate the information from EHRs into the inference of latent representation by making the variational posterior dependent on the associated EHRs during learning. %By conditioning it on the input image $\x$ and its class label $\y$, we can recover the conditional variational auto-encoder (CVAE)~\cite{sohn2015learning}.

\subsection{Knowledge Distillation from EHRs}
\begin{figure}
    \centering
    \includegraphics[width = \textwidth]{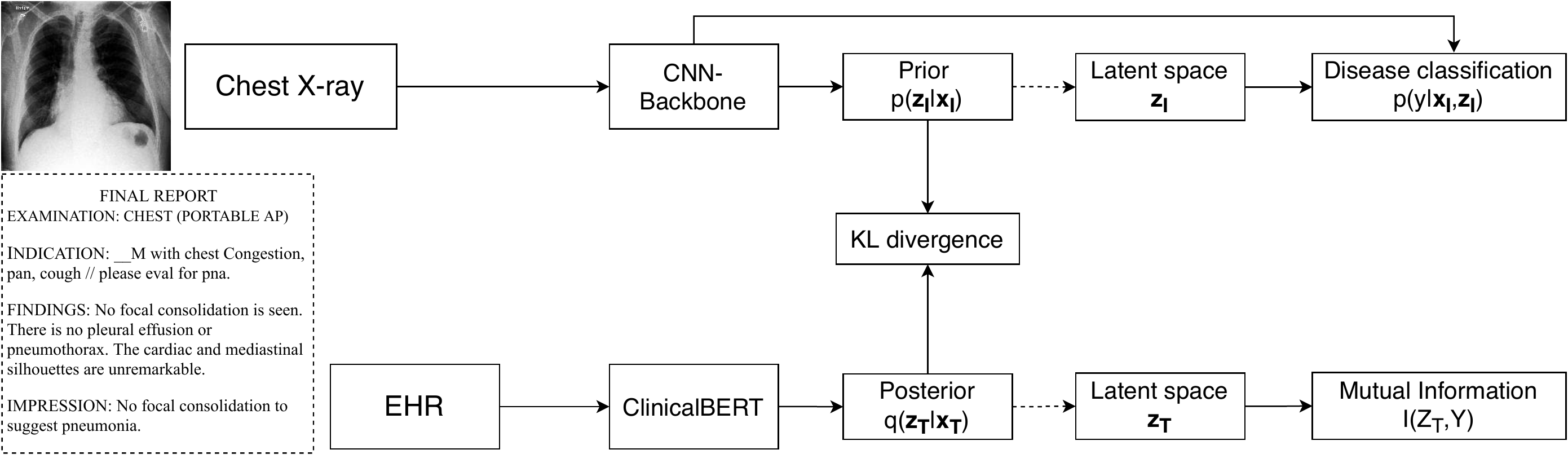} 
    \vspace{-4mm}
    \caption{Illustration of the proposed variational knowledge distillation from EHR texts to X-ray images.}
    \label{fig:overview}
    \vspace{-6mm}
\end{figure}

%We incorporate EHRs into the inference of latent representation of X-ray images so as to improve medical image classification. To this end, w
We introduce a new variational posterior that depends on the corresponding EHR text $\x_T$, which enables us to distill knowledge from EHRs to the representations of images in the latent space. To be more specific, we design the variational posterior as $q(\z_T|\x_T)$, shown in Fig.~\ref{fig:overview}, which gives rise to a new ELBO, as follows:
\begin{equation}
    \mathcal{L}_{\rm{VKD}} = \mathbb{E}\big[\log p(\y|\x_I,\z_I)\big] - D_{\rm{KL}}\big[q(\z_T|\x_T)||p(\z_I|\x_I)\big].
    \label{nelbo}
\end{equation}
By maximizing the above ELBO, the distributional distance in terms of KL divergence between the latent representations of the X-ray image and its associated EHR text is minimized. This encourages the rich knowledge contained in the EHR to be transferred to the image representations. 

In order to extract from the EHR the most relevant information for accurate disease classification, the latent representation $\z_T$ should also be maximally predictive of the disease. This can be achieved by maximizing the mutual information $I(Z_T,Y)$ between $Z_T$ and $Y$, which is intractable. Instead, we can maximize its variational lower bound inspired by \cite{alemi2016deep}, as follows:
\begin{equation}
    I(Z_T,Y) \geq \int p(\x_T) p(\y|\x_T) p(\z_T|\x_T)\log q(\y|\z_T) d\x_T d\y d\z_T = \mathcal{L}_{\rm{MI}},
    \label{mi}
\end{equation} 
where $q(\y|\z_T)$ is the variational approximation of the true predictive distribution $p(\y|\z_T)$. Likewise, we can calculate the empirical approximation of the term on the right hand side in Eq.~(\ref{mi}) by following \cite{alemi2016deep}:
\begin{equation}
    \mathcal{L}_{\rm{MI}} \approx \frac{1}{N} \sum^{N}_{n=1} \int p(\z_T|\x^n_T)\log q(\y^n|\z_T)d\z_T,
    \label{emi}
\end{equation}
where $n$ is the number of the X-ray image and EHR text pairs. In practice, $\mathcal{L}_{\rm{MI}}$ is implemented as a cross entropy loss.

\vspace{-3mm}
\subsection{Empirical Objective Function}

By combining Eqs. (\ref{nelbo}) and (\ref{mi}), we obtain the following empirical objective function for optimization:
% \begin{equation}
%     \mathcal{L} = \mathbb{E}_{p(\z_I|\x_I)}\big[\log p(\y|\x_I,\z_I)\big] + \mathbb{E}_{q(\z_T|\x_T)}\big[\log p(\y|\z_T)\big] - D_{\rm{KL}}\big[q(\z_T|\x_T)||p(\z_I|\x_I)\big]
% \end{equation}
\begin{equation}
\begin{aligned}
    \tilde{\mathcal{L}}_{\rm{VKD}} = &-\frac{1}{N} \sum^{N}_{n=1} \Big[\frac{1}{M}\sum^M_{m = 1} \log p(\y^n|\x_I,\z^{(m)}_I) - \frac{1}{L}\sum^L_{\ell = 1} \log q(\y^n|\z^{(\ell)}_T)\\&+D_{\rm{KL}}\big[q(\z_T|\x^n_T)||p(\z_I|\x^n_I)\big]\Big],
    \label{eq;empobj}
    \end{aligned}
\end{equation}
where $\z^{(m)}_I \sim p(\z_I|\x_I)$, $\z^{(l)}_T \sim q(\z_T|\x_T)$, and $L$ and $M$ are the number of Monte Carlo samples. Note that we take the variational posterior $q(\z_T|\x_T)$ in Eq.~(\ref{nelbo}) as the posterior $p(\z_T|\x_T)$ in Eq.~(\ref{emi}). The resultant objective combines the strengths of the conditional variational auto-encoder and the variational information bottleneck, resulting in a new variational objective for knowledge distribution from EHR texts to X-rays for disease classification. Optimization with $\tilde{\mathcal{L}}_{\rm{VKD}}$ is a process through which the model learns to read X-ray scans like a radiologist to find the relevant visual features to diseases.

%editing link:
%https://drive.google.com/file/d/1JqjPX8ZgRneonynsI5IzZdtO7_M4Z6RY/view?usp=sharing

\subsection{Implementation with Neural Networks}

We implement the optimization objective with deep neural networks (see Fig.~\ref{fig:overview}) by adopting the amortization technique \cite{kingma2013auto}. Both the variational posterior $q(\z_T|\x_T)$ and the prior $p(\z_I|\x_I)$ are parameterized as diagonal Gaussian distributions. To enable back propagation, we adopt the reparameterization trick~\cite{kingma2013auto} for sampling $\mathbf{z}$: $\z^{(\ell)} = f(\x,\epsilon^{(\ell)})$ with $\epsilon^{(\ell)} \sim \mathcal{N}(0,I)$, where $f(\cdot)$ is a deterministic differentiable function. 

 In prior  $p(\z_I|\x_I)$, $\x_I$ is taken as the representation of the X-ray image from a CNN. The inference network of the distribution parameters is implemented by a multi-layer perceptron (MLP). $\x_T$ in the variational posterior $q(\z_T|\x_T)$ is generated through the use of deep contextualized word embeddings. The successful BERT \cite{devlin2018bert} language model based on the Transformer\cite{vaswani2017attention} is used. More specifically, we use a pre-trained version of this model fine-tuned on a database with over two million EHRs~\cite{johnson2016mimic}: ClinicalBERT \cite{huang2019clinicalbert}. To avoid computationally costly fine-tuning, the weights of ClinicalBERT are frozen and a brief fine-tuning step is applied by passing the embeddings through a single trainable Transformer encoder block ($\sim1/12$ the size of ClinicalBERT). Similar to $p(\z_I|\x_I)$, the posterior $q(\z_T|\x_T)$ is also generated by an MLP.

Algorithm~\ref{alg:train} demonstrates the learning process, when the model requires a multi-modal input: an image and the associated EHR. Once trained, the model performs disease classification based on new X-ray scans without the need of an EHR input at inference time, as can be seen in Algorithm~\ref{alg:test}.

\vspace{2mm}
\begin{algorithm}[H]
\KwInput{Training data: $(\x^n_I, \x^n_T, \y^n)$, $n = 1,...,N,$}
\KwOutput{Latent space distributions $p(\z_I|\x_)$, $q(\z_T|\x_T)$, }
\While{not converged:}{
Draw Monte Carlo samples $\z_I^{(m)}$ and $\z_T^{(\ell)}$ from $p(\z_I|\x_I^n)$ and $q(\z_T|\x_T^n)$, respectively\\
Estimate the prediction distributions: $p(\y|\x_I^n,\z_I^{(m)})$ and $q(\y|\z_T^{(\ell)})$\\
Compute $\tilde{\mathcal{L}}_{\rm{VKD}}$ in Eq.~(\ref{eq;empobj})\\
Update models weights via gradient descent on $\tilde{\mathcal{L}}_{\rm{VKD}}$
}
\caption{Learning}
\label{alg:train}
\end{algorithm}

\begin{algorithm}
\KwInput{Testing data with only the X-Ray images $\x_I^j$ .}
\KwOutput{Prediction class label $\y$} 

Draw Monte Carlo samples $\z^{(\ell)}_I$ from the conditional prior $p(\z_I|\x^j_I)$\\
Estimate the prediction distribution $p(\y|\x^j_I,\z^{(\ell)}_I)$\\

\caption{Inference}
\label{alg:test}
\end{algorithm}
% \vspace{-4mm}

\section{Experiments}
\subsection{Datasets}
Three public chest X-ray datasets are used: 1) \textbf{MIMIC-CXR} \cite{johnson2019mimic} is the largest publicly available dataset containing full-text structured EHR and accompanying annotated chest X-rays \cite{johnson2019mimic}. The dataset contains $377,110$ chest x-rays associated with  $227,827$ anonymized EHRs. Each EHR is associated with (multiple) frontal and/or saggital X-ray views, each labelled according to specific classes (e.g. atelectasis, pneumothorax and pneumonia). 2) \textbf{OpenI}\cite{openi} is a similar public dataset with $7,470$ chest X-rays associated with $3,955$ anonymized EHRs. 3) \textbf{Chest X-ray14} \cite{wang2017chest8} contains $112,120$ chest X-rays, without associated EHRs. Paired EHRs exist for this dataset but they are not publicly available. Therefore, we use this dataset for testing but not for training. 

Each image-EHR pair in these datasets is labelled according to a rule-based labelling procedure based on the EHR for fourteen distinct classes. 
MIMIC-CXR is labelled according to a different labeller \cite{irvin2019chexpert} than Chest X-ray14 \cite{wang2017chest8}. These different labelling procedures have an overlap in seven out of fourteen label classes. In this paper the classification labels in MIMIC-CXR are followed.

\subsection{Experimental Settings}

X-ray images are normalized and standardized to grayscale with dimensions of $224\times224$, to align them with the DenseNet-121 CNN backbone, pre-trained on ImageNet \cite{He_2016_CVPR}. Pre-trained CNN backbones have been proven effective in similar medical applications \cite{Raghu2019}, and DenseNet-121 specifically has been proven ideal for X-ray images\cite{rajpurkar2017chexnet,xue2019improved}. 
Each EHR is tokenized according to WordPiece \cite{wu2016google} tokenization, which has a library of around $30 000$ tokens. Each tokenized EHR is preceded by a $[CLS]$ classification token and ended with a $[SEP]$ token, following the methodology used in \cite{devlin2018bert,huang2019clinicalbert}. The maximum number of tokens is set to $256$. Shorter EHRs are zero padded to obtain text embeddings of the same sizes. 
The size of latent spaces $\z_I$ and $\z_T$ is set to an empirically determined value of 512. Two-layer MLPs with layer sizes $\{512, 512\}$ are used for the amortized inference of the prior and variational posterior. A dropout rate of 0.5 is applied to all layers, except the CNN backbone and the final layer in the generation of latent space $\z$, to which no dropout is applied. These architectures are trained on an NVIDIA RTX 2080ti GPU, using Adam\cite{kingma2014adam} optimization for a duration defined by early stopping with a tolerance of 1\%.

A common problem in optimizing variational architectures is KL vanishing. To prevent this, cyclical KL annealing \cite{fu2019cyclical} is applied according to Eq.~\ref{eq:an1}, where the KL loss is multiplied with $\beta_t$. $g(\tau)$ is a monotonically increasing function, T is the number of batch iterations, t the current iteration, R (=0.5) determines the annealing grade and C (=4) is the number of annealing cycles per epoch:

\begin{equation}
    \beta_t = \left\{
    \begin{array}{l}
    g(\tau),\quad \tau \leq R\\
    1,\quad\quad\,\: \tau > R 
    \end{array}
    \right\},
    \quad
    \text{where} ~~~\tau = \frac{\text{mod}(t-1, [T/C])}{T/C} 
    \label{eq:an1}
\end{equation}

\subsection{Results}
\begin{table}[!b]
\centering
\resizebox{\textwidth}{!}{%
\begin{tabular}{|l|ccccc|cccc|cc|}
                            \cline{2-12}
                           \multicolumn{1}{c|}{}& \multicolumn{5}{c|}{Chest X-ray14} & \multicolumn{4}{c|}{Open-I}                                                & \multicolumn{2}{c|}{MIMIC-CXR}                                 \\ \cline{2-12}
                           \multicolumn{1}{c|}{}& \cite{rajpurkar2017chexnet}  & \cite{xue2019improved}    & \cite{wang2017chest8}  &\cite{chen2020label}& Ours  & \cite{wang2017chest8} & \cite{Nunes2019}& \begin{tabular}[c]{@{}c@{}}Ours \\ (no EHR)\end{tabular} & Ours& \begin{tabular}[c]{@{}c@{}}Ours\\ (no EHR)\end{tabular} & Ours\\\hline
No Finding                 &  -        & -      & -&-        &  -    &  -      &-&0.711&     \textbf{0.720}                              &0.825 &\textbf{0.827}     \\ %\hline
\begin{tabular}[c]{@{}l@{}}Enlarged \\ Cardiomediastinum\end{tabular} & - & -& -&-&  -    &      -  & -&- &     -               &  0.589&\textbf{0.838}     \\ %\hline
Cardiomegaly               & 0.889      & 0.892  & 0.810 &0.893  & \textbf{0.899}     &    0.803    &         -&0.837&\textbf{0.851}     &0.739&\textbf{0.758}     \\ %\hline
Lung Opacity               & -          & -      &  -&-       & -     &  -      &-&\textbf{0.720}&     0.698                             &\textbf{0.698}&0.695     \\ %\hline
Lung Lesion                & -          & -      &  -&-       &   -   &  -      &    -&0.539&     \textbf{0.710}                         &0.663&\textbf{0.690}     \\ %\hline
Edema                      & 0.888      & \textbf{0.898}  & 0.805&0.850   &0.893      &   0.799     & -&0.897                   &\textbf{0.923}& 0.832&\textbf{0.861}     \\ %\hline
Consolidation              & 0.790      & 0.813  & 0.703&0.751   & \textbf{0.819}     &    0.790    & -&\textbf{0.859}                   &0.652& 0.731&\textbf{0.783}     \\ %\hline
Pneumonia                  & 0.768      & 0.767  & 0.658&      0.739&   \textbf{0.781}&   \textbf{0.642}     & -&0.610                   &0.619&0.618&\textbf{0.627}     \\ %\hline
Atelectasis                & 0.809      & 0.822  & 0.700&   0.786 &\textbf{0.825}    &    0.702    &  -&0.771&    \textbf{0.797}         & 0.725&\textbf{0.749}     \\ %\hline
Pneumothorax               & 0.889      & 0.870  & 0.799&   0.876&    \textbf{0.903}&    0.631   &-&\textbf{0.784} &     0.637           &0.721&\textbf{0.758}     \\ %\hline
Pleural Effusion           & 0.864      & \textbf{0.881}      & 0.759& 0.832       & 0.871     &   0.890     &   -&\textbf{0.904} &0.858 &0.864&\textbf{0.892}    \\ %\hline
Pleural Other              &-           & -      &  -&-       &  -    &  -      &    -&0.637&     \textbf{0.876}                         &0.731&\textbf{0.776}     \\ %\hline
Fracture                   &         -  & -      &  -&-       &  -    & -       &    -&0.486 &     \textbf{0.532}                        &0.557&\textbf{0.698}     \\ %\hline
Support Devices            &  -         & -      & - &-       &-      & -       &-&0.553 &     \textbf{0.581}                            &0.854&\textbf{0.897}     \\ \hline
Average                    & -          & 0.842  & 0.722   &0.826      &\textbf{0.872}        & 0.719   &0.621&0.837&\textbf{0.885}      &  0.807      &\textbf{0.839}  \\
\hline
\end{tabular}
}
\caption{Comparison of AUC values per class for NIH chest X-ray14 (partial), OpenI and MIMIC-CXR datasets. }
\label{tab:comparison}
\end{table}

\paragraph{\textbf{State-of-the-Art Comparison.}} The performance of our architecture in comparison with earlier works on image-based chest X-ray classification \cite{xue2019improved, rajpurkar2017chexnet, wang2018tienet,wang2017chest8,Nunes2019} is shown in Table \ref{tab:comparison}. We report the results of the proposed method with and without variational knowledge distillation (i.e., no EHR). 
Results of our proposed method on the Chest X-ray14 dataset are obtained by fine-tuning a model pre-trained on MIMIC-CXR. Note that the fine-tuning step is necessary to alleviate domain shift between different datasets. Results on the OpenI and MIMIC-CXR datasets are obtained without any specific pre-training on radiology images from other datasets. 

Results on the OpenI and MIMIC-CXR datasets show the performance gain due to knowledge distillation, where the performance improvement is consistent on the latter vastly larger dataset. It is worth mentioning that the high performance on Chest X-ray14 further indicates that the proposed variational knowledge distillation is transferable between datasets, even when the new target dataset does not contain EHRs. Note that our approach outperforms all previous approaches for chest X-ray classification.

\paragraph{\textbf{Ablation Studies.}} To verify the effectiveness of all elements in the proposed architecture two ablation studies are conducted. In the first ablation study, $\mathcal{L}_{\rm{MI}}$ (Eq.~(\ref{emi})) is left out of the objective function, thus there is no specific requirement for $\z_T$. Moreover, only the classification objective from the image branch is taken into account, removing the KL term, which results in a regular image-based classifier. This first ablation study (Table~\ref{tab:ab1}) reveals that the major contributing factor to the performance of our method shows to be variational knowledge distillation, whereas the addition of the objective function in the EHR branch ($\mathcal{L}_{\rm{MI}}$) has a relatively smaller, yet considerable effect. 
Secondly, we test the effect of the size of latent space $\z$ on the performance (Table~\ref{tab:ab2}). It appears that increasing the size of $\z$ can improve model performance, while this effect tends to be smaller with larger values of $\z$. The current value of $\z$ was chosen to be $512$, which maximizes performance against computational cost.
%Increasing the size of $\z$ even further might offer small performance gains . 

\begin{table}[!t]
\begin{minipage}[b]{.50\textwidth}
   \centering

\begin{tabular}[]{|l|c|c|}
\hline
                                          & OpenI     & MIMIC-CXR           \\ \hline
Full architecture                              & \textbf{0.885}           & \textbf{0.839}                             \\ \hline
w/o $\mathcal{L}_{\rm{MI}}$ (Eq.~\ref{emi})               & 0.873          & 0.832                          \\ \hline
w/o VKD (no EHR)                   &  0.837         &  0.807                  \\\hline        
\end{tabular}

   \caption{AUC scores for varying architecture compositions.}
   \label{tab:ab1}
\end{minipage}\qquad
\begin{minipage}[b]{.50\textwidth}
   \begin{tabular}[]{|l|c|c|}
      \hline
                                              Size of $\mathbf{z}$ & OpenI     & MIMIC-CXR           \\ \hline
    64                               & 0.814           & 0.775                             \\ \hline
    128                & 0.870          & 0.829                          \\ \hline
    512 - default                   &  0.885         &  0.839    \\\hline
    1024                  &  \textbf{0.891}         & \textbf{0.842}   \\\hline 
    \end{tabular}
    \captionsetup{width=\linewidth}
   \caption{AUC scores for varying \\size of latent space $\z$.}
   \label{tab:ab2}
\end{minipage}
\vspace{-10mm}
\end{table}

\begin{figure}[!b]
  \setlength{\fboxsep}{0pt}%
  \setlength{\fboxrule}{0.3pt}%
  \begin{subfigure}[b]{0.39\textwidth}
    \fbox{\includegraphics[width=\textwidth-1pt]{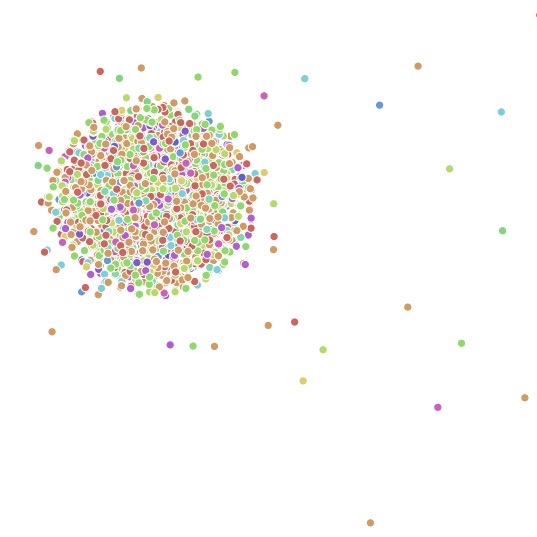}}
    \caption{Without VKD}
    \label{fig:f1}
  \end{subfigure}
  \hfill
  \begin{subfigure}[b]{0.39\textwidth}
    \fbox{\includegraphics[width=\textwidth-1pt]{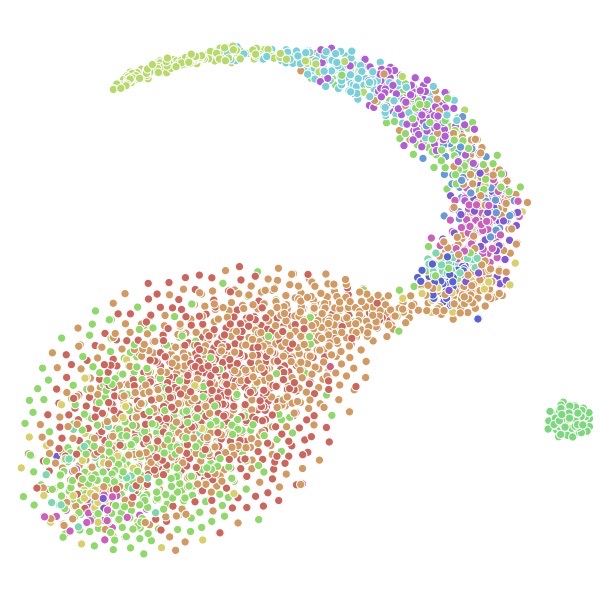}}
    \caption{With VKD}
    \label{fig:f2}
  \end{subfigure}
  \hfill
  \begin{subfigure}[t]{0.18\textwidth}
    \raisebox{10mm}{\includegraphics[width=\textwidth]{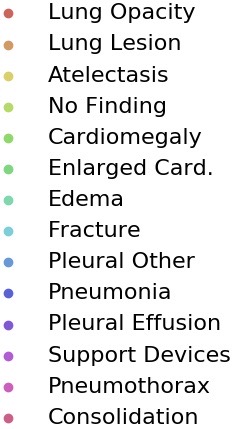}}

  \end{subfigure}
  \caption{t-SNE embeddings of latent space $\z_I$ with and without VKD, overlayed with class labels, showing that VKD causes structuring of $\z_I$.}
  \label{fig:tsne}
\end{figure}

% \begin{figure}
%     \centering
%     \includegraphics[width = \textwidth]{llncs2e/latent_space_umap.png}
%     \caption{U-map embeddings of latent space from prior network, overlayed with class labels for each data point. a) without knowledge distillation from EHR and b) with the full architecture.  \tom{legend tb fixed} }
%     \label{fig:lsviz}
% \end{figure}

\paragraph{\textbf{Visualizations.}}

\begin{figure}[!t]
\begin{subfigure}{.5\textwidth-1.05pt}
\centering
\includegraphics[width=.83\linewidth]{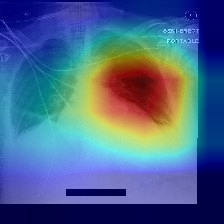}

\end{subfigure}
\hfill
\begin{subfigure}{.5\textwidth-1.05pt}
\centering
\includegraphics[width=.83\linewidth]{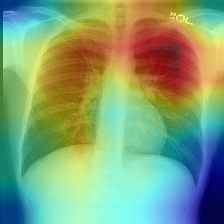}

\end{subfigure}
\setlength{\fboxsep}{0pt}%
\setlength{\fboxrule}{0.3pt}%
\begin{subfigure}{.5\textwidth-1.05pt}

\centering
\fbox{\includegraphics[width=.83\linewidth]{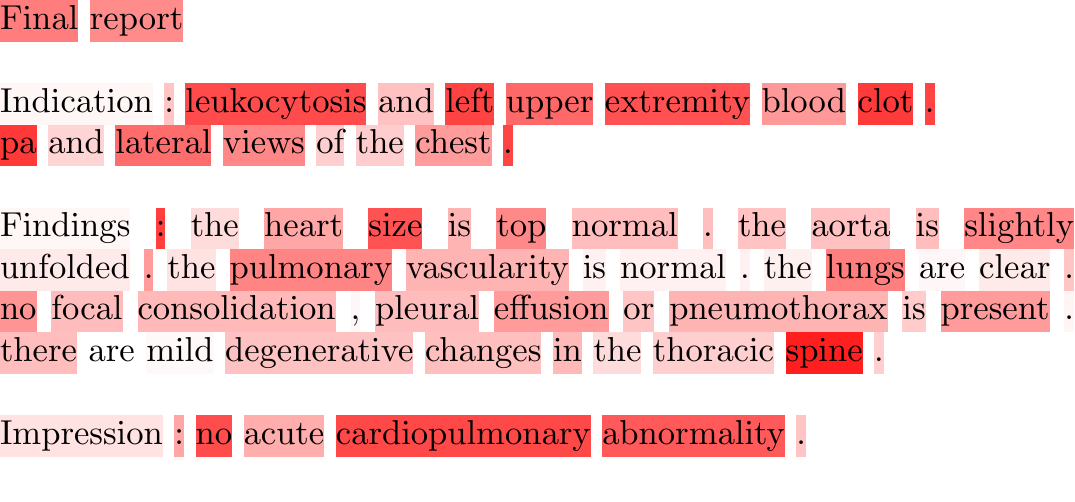}}
\end{subfigure}
\hfill
\begin{subfigure}{.5\textwidth-1.05pt}
\centering
\fbox{\includegraphics[width=.83\linewidth]{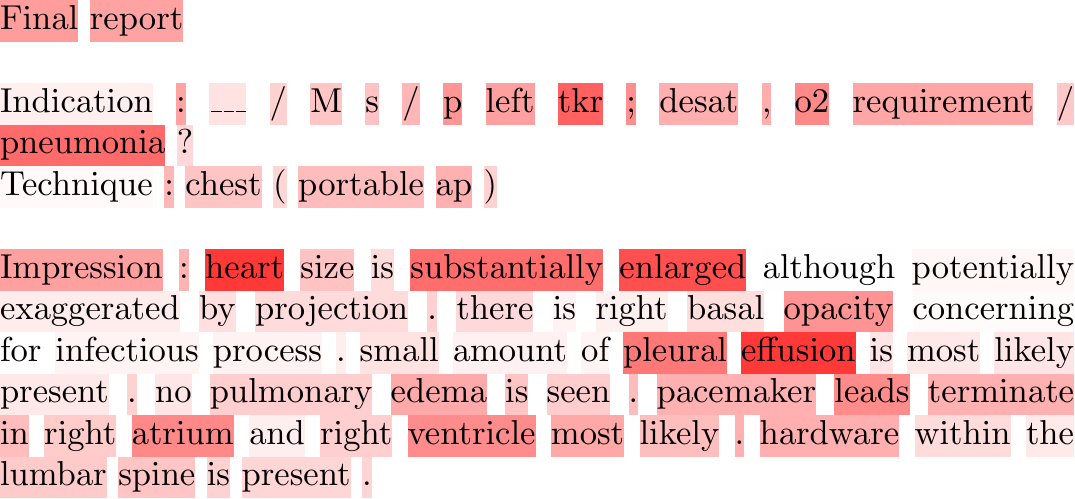}}

\end{subfigure}
\caption{Importance of words and image regions for two image-EHR pairs. WordPiece tokens of the EHRs are averaged if needed to form full words. Darker red means higher importance.}
\label{fig:mmweights}
\vspace{-6mm}
\end{figure}
The difference between $\z_I$ with and without variational knowledge distillation is shown in Fig.~\ref{fig:tsne}. With variational knowledge distillation a structurenedness in classes within $\z_I$ can be observed, whereas without it there seems to be more reliance on image tokens directly passed to the classification head, consequently resulting in a less structured $\z_I$.   
In Fig.~\ref{fig:mmweights} weight visualizations for the final CNN-layer with Grad-cam\cite{selvaraju2017grad} and the final Transformer layer with BertViz\cite{vig2019multiscale} are shown for images and EHRs respectively. As can clearly be seen the visual focus is correctly on the lung region. Weights of the EHR tokens show a clear emphasis on important nouns and adjectives in the EHR. Verbs and prepositions show lower weights. These visualizations provide an intuitive illustration that our model is able to extract visual features relevant to the disease due to the proposed variational knowledge distillation.

\section{Conclusion}

In this paper, we propose a new probabilistic inference framework of multi-modal learning for disease classification based on X-rays by leveraging EHRs. 
We developed a latent variable model to learn latent representations of X-ray images. We introduce variational knowledge distillation that enables the model to acquire the ability to extract visual features relevant to the disease. This strategy enables us to incorporate the knowledge in EHRs during training, without relying on them in the testing stage. We conduct experiments on the current largest and most widely used chest X-ray - EHR datasets: MIMIC-CXR and OpenI, showing the benefit of variational knowledge distillation. Moreover we demonstrate our method performs well on Chest X-ray14 with only images by pre-training on MIMIC-CXR, which indicates its 
strong transfer ability across datasets. 

%More importantly, it opens possibilities for supervised learning from large annotated databases, without being dependent on them during clinical application.  

\bibliography{references}
\bibliographystyle{splncs04}
\end{document}